\documentclass[a4paper,11pt]{article}
\usepackage{pos}
\usepackage{bm}
\usepackage{tikz}
\usepackage{comment}
\usepackage{color,xcolor}
\usepackage[normalem]{ulem}
\usetikzlibrary{arrows.meta,calc,decorations.pathmorphing,positioning}

\newcommand{\Q}{\mathcal Q}
\newcommand{\pt}{p_{T}}

\newcommand{\as}{\alpha_s}
\newcommand{\MSbar}{\overline{\mathrm{MS}}}

\title{Towards High-Energy Factorization at Full NLO for Quarkonium Production}
\ShortTitle{Towards High-Energy Factorization at Full NLO for Quarkonium Production}

\author*[a]{Michael Fucilla}
\author[b]{Jean-Philippe Lansberg}
\author[c]{Maxim Nefedov}
\author[a]{Lech Szymanowski}
\author[b]{Samuel Wallon}

\affiliation[a]{National Centre for Nuclear Research,\newline
  Pasteura 7, 02-093 Warsaw, Poland}
\affiliation[b]{Universit\'e Paris-Saclay, CNRS/IN2P3, IJCLab,\newline
  F-91405 Orsay, France}
\affiliation[c]{Physics Department, Ben-Gurion University of the Negev,\newline
  Beer Sheva 84105, Israel}

\emailAdd{Michael.Fucilla@ncbj.gov.pl}

\abstract{Heavy-quarkonium hadroproduction at the LHC can access kinematic configurations in which the partonic center-of-mass energy is much larger than the quarkonium mass and transverse momentum. Fixed-order predictions may become unstable in this regime because large
logarithms of the collision energy are not resummed. Extending high-energy
factorization beyond leading logarithms requires process-dependent impact
factors at next-to-leading order (NLO). We report the first complete NLO BFKL
impact factors, with exact heavy-quark-mass dependence, for the production of
the NRQCD states ${}^1S_0^{[1]}$, ${}^1S_0^{[8]}$, and ${}^3S_1^{[8]}$. The
real-emission calculation is combined with the previously known virtual
corrections through a local subtraction procedure. Soft singularities cancel,
initial-state collinear poles are absorbed into the parton distributions, and
rapidity divergences are removed by BFKL factorization. These results provide
the ingredients needed for full-NLL studies of forward quarkonium and of
quarkonium-associated production with a large rapidity separation.}

\FullConference{The 33rd International Workshop on Deep Inelastic Scattering and Related Subjects (DIS2026)\\
4--8 May 2026\\
Bologna, Italy}

\begin{document}
\maketitle

\section{Motivation}

Heavy quarkonia offer a distinctive testing ground for perturbative QCD. The
heavy-quark mass supplies a hard scale even when the observed transverse
momentum is moderate, while the transition of the produced heavy-quark pair
into a physical bound state can be organized within non-relativistic QCD
(NRQCD)~\cite{Bodwin:1994jh,Lansberg:2019adr}. At hadron colliders, however,
the available partonic energy can be much larger than both the quarkonium mass
and its transverse momentum. The perturbative series then contains powers of large logarithms of the ratio
between the center-of-mass energy, $\sqrt{\hat{s}}$, and a semi-hard scale, $Q$,
\begin{equation}
 \as^n\ln^n\!\frac{\hat s}{Q^2}
 \quad\text{and}\quad
 \as^{n+1}\ln^n\!\frac{\hat s}{Q^2},
 \qquad \hat s\gg Q^2\gg\Lambda_{\rm QCD}^2,
 \label{eq:logs}
\end{equation}
which are resummed at leading-logarithmic (LL) and next-to-leading-logarithmic
(NLL) accuracy, respectively, within the BFKL approach~\cite{Kuraev:1976ge,
Kuraev:1977fs,Balitsky:1978ic,Fadin:1998py,Ciafaloni:1998gs}; for pioneering full NLL studies see e.g.~\cite{Colferai:2010wu,Ducloue:2013bva,Caporale:2014gpa}. Systematic matching of the BFKL resummation with collinear factorisation in the $\overline{\rm MS}$-scheme is achieved through the High-Energy Factorisation formalism~\cite{Catani:1990eg,Catani:1994sq}.

The resummation is not merely a formal refinement. Inclusive quarkonium
cross sections computed at fixed NLO can develop a strong scale dependence or
even become negative as the energy increases~\cite{Schuler:1994hy,Mangano:1996kg,Feng:2015cba}, as a consequence of the hierarchy $\sqrt{s} \gg M = 2 m_Q$. This behavior has been recently studied in details,
for example, in inclusive vector-quarkonium photoproduction~\cite{ColpaniSerri:2021bla} and
pseudoscalar-quarkonium hadroproduction~\cite{Lansberg:2020ejc}. Matching fixed-order calculations to
high-energy resummation restores a stable energy dependence already at
LL accuracy~\cite{Lansberg:2021vie,Lansberg:2023kzf,Flett:2024htj}. Quantitative predictions
at small $x$, however, call for the next logarithmic order.

In high-energy factorization, the process dependence is contained in impact
factors, whereas radiation spanning the large rapidity interval is described
by a universal BFKL Green's function. NLL accuracy therefore requires both the
NLL BFKL kernel and NLO impact factors. The former has long been
known~\cite{Fadin:1998py,Ciafaloni:1998gs}, while a few impact factors are known at full NLO accuracy~\cite{Bartels:2001ge,Bartels:2002yj,Caporale:2011cc,Ivanov:2004pp,Polizzi:2025edm,Nefedov:2019mrg,Hentschinski:2020tbi,Celiberto:2022fgx,Fucilla:2024cpf}. Massive quarkonium impact factors were available only at leading order~\cite{Kniehl:2006sk}. NLO treatments in
BFKL phenomenology instead relied on fragmentation, and hence on the
leading-power limit $\pt^2\gg m_Q^2$~\cite{Celiberto:2022dyf}, with $\pt$ the transverse momentum of the quarkonium and $m_Q$ the mass of the heavy-quark constituent. The calculation
summarized here removes this restriction for three phenomenologically
important $S$-wave NRQCD states~\cite{Fucilla:2026nlo}.

\section{Quarkonium production in the BFKL--NRQCD framework}

Consider two tagged objects produced far apart in rapidity, for instance a
quarkonium pair,
\begin{equation}
 h(P_1)+h(P_2)\longrightarrow \Q(p_1)+X+\Q(p_2).
 \label{eq:process}
\end{equation}
The symbol $X$ denotes the inclusive radiation between the tags. After the
standard convolution with collinear parton distributions, the resummed
partonic cross section has the schematic form
\begin{equation}
 d \hat{\sigma}_{ab}=
 \int\!\frac{d^2\bm q_1}{\bm q_1^2}
       \frac{d^2\bm q_2}{\bm q_2^2}\,
 V_a^{(\Q_1)}(\bm q_1,x_1)\,
 G_{\rm BFKL}(\bm q_1,\bm q_2;Y)\,
 V_b^{(\Q_2)}(\bm q_2,x_2),
 \label{eq:bfklfact}
\end{equation}
where $Y$ is the rapidity interval, $G_{\rm BFKL}$ is universal, and
$V_{a,b}$ describes the transition from an incoming parton to the corresponding
tagged final state. 

The same ingredients enter several high-energy configurations. When both
objects are tagged in opposite fragmentation regions, the large logarithm is
generated by their rapidity separation and the Green's function connects two
perturbative impact factors. For single-forward production, one of these
vertices is instead combined with a proton impact factor and the BFKL Green function,
or equivalently, with what is known as an unintegrated gluon distribution. Central production
requires evolution on both sides of a central-emission vertex. The NLO
quarkonium vertex discussed below is therefore relevant to the first two
configurations but not for the third.

NRQCD further factorizes each quarkonium impact factor into perturbative
short-distance coefficients and long-distance matrix elements (LDMEs),
\begin{equation}
 V_a^{(\Q)}=\sum_m
 V_a^{(m)}\,\langle\mathcal O^{\Q}(m)\rangle,
 \qquad m={}^{2S+1}L_J^{[1,8]}.
 \label{eq:nrqcd}
\end{equation}
The label $m$ specifies the spin, orbital angular momentum, total angular
momentum, and color representation of the intermediate $Q\bar Q$ pair. We
considered the color-singlet state ${}^1S_0^{[1]}$ and the color-octet states
${}^1S_0^{[8]}$ and ${}^3S_1^{[8]}$. They contribute, in particular, to
$\eta_c$ and $\eta_b$ production and to the color-octet sector of $J/\psi$ and
$\Upsilon$ production. Their role in the NRQCD expansion is summarized in
Table~\ref{tab:states}. At Born level the relevant transition is
\begin{equation}
 g + R(q)\longrightarrow Q\bar Q[m](p),
 \label{eq:born}
\end{equation}
where $R$ is a reggeized gluon carrying transverse momentum. Its impact factor
is 
\begin{equation}
 V_g^{(m,0)}(\bm q,\bm p,z)=
 h_m^{(0)}(\bm p^2)\,
 \delta^{(2)}(\bm q-\bm p)\,\delta(1-z) \; ,
 \label{eq:bornif}
\end{equation}
where $z$ is the fraction of longitudinal momenta of the incoming parton carried by the quarkonium, while $\boldsymbol{q}$ and $\boldsymbol{p}$ are the Reggeized gluon and quarkonium transverse momentum, respectively. The functions $h_m^{(0)}$ retain the complete dependence on
$M\simeq2m_Q$ and $\bm p^2$. In particular, no expansion in
$m_Q^2/\pt^2$ is made.

\begin{table}[t]
\centering
\small
\begin{tabular}{ccl}
\hline
NRQCD state & color & representative physical contribution \\
\hline
${}^1S_0^{[1]}$ & singlet & leading channel for $\eta_c$ and $\eta_b$ \\
${}^1S_0^{[8]}$ & octet & $J/\psi$, $\Upsilon$, $\eta_c$, and $\eta_b$ \\
${}^3S_1^{[8]}$ & octet & $J/\psi$, $\Upsilon$, $\eta_c$, and $\eta_b$ \\
\hline
\end{tabular}
\caption{The $S$-wave intermediate states whose massive impact factors are
known at NLO. Their weights in a physical cross section are set by the
corresponding LDMEs.}
\label{tab:states}
\end{table}

\section{The massive impact factor at NLO}

The virtual corrections to Eq.~\eqref{eq:born} were obtained using Lipatov's
gauge-invariant effective action~\cite{Lipatov:1995pn} and are known for all
three states~\cite{Nefedov:2024swu}. Completing the NLO impact factor requires
the real subprocesses
\begin{align}
 g+R&\longrightarrow Q\bar Q[m]+g,
 &q+R&\longrightarrow Q\bar Q[m]+q,
 \label{eq:realchannels}
\end{align}
including both ordinary QCD diagrams and the induced reggeon--gluon couplings
that restore gauge invariance for the off-shell $t$-channel exchange. The
quark-initiated channel first appears at this order.

The amplitudes were evaluated directly in the high-energy effective theory,
with covariant NRQCD projectors inserted before squaring. The induced
interactions are essential: the subset obtained by replacing an external
on-shell gluon with an off-shell reggeized gluon is not gauge invariant by
itself. The complete result passes the corresponding Ward-identity tests and
retains all powers of $M^2/\pt^2$. This last feature distinguishes the
calculation from NLO fragmentation vertices, whose domain is restricted to
large transverse momentum.

The exact squared matrix elements depend on $z$, the transverse momenta, and
the heavy-quark mass. Their analytic expressions are lengthy, but their
singular limits have a compact universal organization. Denoting
$\bm k=\bm q-\bm p$, the real contribution can be separated as
\begin{align}
 V_{a,\mathrm{fin}}^{(m,1)}(\bm q,\bm p,z)
 ={}& \mathcal N_m(\bm p,z)
 \left[
 \widetilde H_{Ra}^{(m)}(\bm q,\bm p,z)
 -\mathcal J_{Ra}^{(m)}(\bm k,\bm p,z)
 \right]_{\epsilon=0},
 \label{eq:subtraction}
\end{align}
where $\widetilde H_{Ra}^{(m)}$ is the exact reduced squared matrix element.
The subtraction term $\mathcal J_{Ra}^{(m)}$ reproduces every non-integrable
soft, collinear, and rapidity limit. Consequently, the difference in brackets
can be integrated numerically in four dimensions, whereas
$\mathcal J_{Ra}^{(m)}$ can be integrated analytically in
$d=4-2\epsilon$ dimensions.

For the gluon channel, the singular content of the subtraction term can be
displayed schematically as
\begin{align}
 \mathcal J_{Rg}^{(m)}={}&
 \frac{2C_A}{\bm k^2}
 \left[
  \frac{1-z}{(1-z)^2+r\bm k^2/(q^+)^2}
  +\frac{1}{z}+z(1-z)-2
 \right] +\Delta\mathcal J_{Rg}^{(m,\mathrm{pol})}
  +\Delta\mathcal J_{Rg}^{(m,\mathrm{CO})},
 \label{eq:Jg}
\end{align}
up to a multiplicative global Born normalization. The first line contains the standard
initial-state singularities and rapidity divergences. The polarization-dependent term is needed for
the spin-singlet channels, while the last term occurs only when the produced
pair carries a net color charge. It subtracts the additional soft radiation
from a color-octet final state. The quark channel contains the corresponding
$q\to qg$ splitting structure but no rapidity pole for $z \rightarrow 1$.

The parameter $r$ in Eq.~\eqref{eq:Jg} implements tilted-Wilson-line
regularization in the high-energy effective action. The distributional
identity
\begin{equation}
 \mathcal J_{RR}^{(r)}(\bm k) =\frac{1-z}{(1-z)^2+r\bm k^2/(q^+)^2}
 =\frac{1}{(1-z)_+}
 -\frac{1}{2}\delta(1-z)
  \ln\!\frac{r\bm k^2}{(q^+)^2}+\mathcal O(r)
 \label{eq:rapidity}
\end{equation}
makes the rapidity logarithm explicit. Its coefficient is fixed by the LL
BFKL kernel. This provides a stringent check of high-energy factorization and
allows the result to be translated between the effective-action regulator,
the conventional BFKL cutoff, and other rapidity-factorization schemes.
For example, the change from the tilted-Wilson-line prescription to a cutoff
$s_\Lambda$ can be represented by the finite difference
\begin{align}
 \Delta\mathcal J_{RR}(\bm k) = \frac{2C_A}{\bm k^2}
 \left[
 \frac{\Theta\!\left(1-z-\bm k^2/s_\Lambda\right)}{1-z}
 -\frac{1-z}{(1-z)^2+r\bm k^2/(q^+)^2}
 \right].
 \label{eq:schemechange}
\end{align}
The impact factor and the Green's function change separately under this
redefinition, while their convolution is invariant to NLL accuracy. This
scheme conversion is needed before inserting the result into a numerical BFKL
implementation that uses a different rapidity cutoff.

\section{Cancellation and factorization of singularities}

After analytic integration of the subtraction terms, the NLO impact factor
can be organized as
\begin{equation}
 V_a^{(m,1)}=
 V_{a,\mathrm{fin}}^{(m,1)}+
 V_{a,\mathrm{sub}}^{(m,1)}+
 V_{a,\mathrm{virt}}^{(m,1)}+
 V_{a,\mathrm{PDF}}^{(m,1)}+
 V_{a,\mathrm{BFKL}}^{(m,1)}.
 \label{eq:decomposition}
\end{equation}
Each term in Eq.~\eqref{eq:decomposition} may separately contain poles, but
their sum is finite. The cancellation takes place in four conceptually
distinct steps.

First, ultraviolet poles are removed by $\MSbar$ renormalization of the strong
coupling and on-shell renormalization of the heavy-quark mass and external
fields. Second, the soft poles cancel out between real and virtual emission, as
required by the KLN mechanism. In the color-octet channels this includes the
extra soft singularity associated with radiation from the final-state
$Q\bar Q$ pair. Its cancellation is an important check that is absent in the
color-singlet case.

Third, the remaining initial-state collinear poles are proportional to the
usual splitting functions. They are absorbed into the renormalized collinear
PDFs,
\begin{equation}
 -\frac{1}{\epsilon}P_{ab}(z)
 \quad\longrightarrow\quad
 P_{ab}(z)\ln\!\frac{\bm p^2}{\mu_F^2},
 \label{eq:collinear}
\end{equation}
leaving the expected dependence on the factorization scale $\mu_F$. Finally,
the rapidity divergence in Eq.~\eqref{eq:rapidity} is removed by the BFKL
counterterm and assigned to the evolution of the Green's function. The
resulting scale dependence cancels, to the calculated accuracy, against that
of NLL BFKL evolution.

These cancellations were verified independently for the gluon- and
quark-induced channels and for each of the states
${}^1S_0^{[1]}$, ${}^1S_0^{[8]}$, and ${}^3S_1^{[8]}$. The final answer is a
sum of compact analytic distributions and a finite remainder retaining the
full $m_Q^2/\pt^2$ dependence; the complete expressions and computer-readable
matrix elements are given in Ref.~\cite{Fucilla:2026nlo}. At large transverse
momentum, logarithms of $\pt^2/m_Q^2$ emerge and connect the massive result to
the fragmentation regime. At $\pt\sim m_Q$, where a leading-power expansion
is not justified, the same result remains directly applicable.

\section{Phenomenological opportunities}

The immediate application is associated production with a large rapidity
separation, such as $J/\psi$ plus a backward jet or two tagged quarkonia. These
channels are quarkonium analogues of Mueller--Navelet jets~\cite{Mueller:1986ey}
and have so far been studied with massive quarkonium vertices only at
LL~\cite{Boussarie:2017oae,He:2019qqq}. Combining the impact factors presented
here with the NLL Green's function makes a consistent full-NLL analysis
possible.

The exact mass dependence is especially relevant at moderate transverse
momentum. For a fixed rapidity interval, large transverse masses require
larger incoming momentum fractions and are therefore increasingly suppressed
by the PDFs. Moderate-$\pt$ quarkonia can probe the same rapidity interval at
smaller $x$, where the cross section is larger and high-energy evolution has
more room to develop. This is precisely the region that cannot be described
reliably by a fragmentation-only impact factor~\cite{Celiberto:2022fgx}.

A second application is forward single-quarkonium production within
high-energy factorization~\cite{Catani:1990eg,Catani:1994sq}. The NLO vertex can be
convoluted with an evolved unintegrated gluon distribution and matched to the
fixed-order collinear result. Such a calculation would extend the existing LL
matching studies~\cite{Lansberg:2021vie,Lansberg:2023kzf} to full NLL
accuracy. The present states supply the color-singlet contribution to
pseudoscalar quarkonia and important color-octet contributions to vector
quarkonia. Completing the full NRQCD prediction at the same perturbative
order will additionally require the NLO impact factors for the relevant
$P$-wave channels.

\section{Conclusions}

We have reported the first complete NLO BFKL impact factors for inclusive
production of the NRQCD states ${}^1S_0^{[1]}$, ${}^1S_0^{[8]}$, and
${}^3S_1^{[8]}$, keeping the heavy-quark mass exactly. The
result exhibits the expected cancellation of soft poles, collinear
factorization into the PDFs, and high-energy factorization of the rapidity
divergence. It supplies the last missing ingredient for NLL studies of quarkonium production within the BFKL approach

\paragraph{Acknowledgments.}
M.N. is supported by BSF grants 2022132 and 2021789, ISF grant 910/23, and the
MSCA project \emph{RadCor4HEF} (grant 101065263). M.F. is supported by the
NAWA Ulam fellowship BNI/ULM/2024/1/00065. The work also received support from
the P2I Graduate School of Physics, the EU \emph{Strong2020} project, the
French ANR project ANR-20-CE31-0015 (\emph{PrecisOnium}), and the CNRS--IN2P3
project \emph{QCDFactorisation@NLO}. L.S. is supported by the Polish National
Science Centre under grant 2024/53/B/ST2/00968.


\begin{thebibliography}{99}

\bibitem{Bodwin:1994jh}
G.~T. Bodwin, E.~Braaten and G.~P. Lepage,
Phys. Rev. D \textbf{51} (1995) 1125

\bibitem{Lansberg:2019adr}
J.-P. Lansberg,
Phys. Rept. \textbf{889} (2020) 1

\bibitem{Kuraev:1976ge}
E.~A. Kuraev, L.~N. Lipatov and V.~S. Fadin,
Sov. Phys. JETP \textbf{44} (1976) 443.

\bibitem{Kuraev:1977fs}
E.~A. Kuraev, L.~N. Lipatov and V.~S. Fadin,
Sov. Phys. JETP \textbf{45} (1977) 199.

\bibitem{Balitsky:1978ic}
I.~I. Balitsky and L.~N. Lipatov,
Sov. J. Nucl. Phys. \textbf{28} (1978) 822.

\bibitem{Fadin:1998py}
V.~S. Fadin and L.~N. Lipatov,
Phys. Lett. B \textbf{429} (1998) 127

\bibitem{Ciafaloni:1998gs}
M.~Ciafaloni and G.~Camici,
Phys. Lett. B \textbf{430} (1998) 349

\bibitem{Colferai:2010wu}
D.~Colferai, F.~Schwennsen, L.~Szymanowski and S.~Wallon,
JHEP \textbf{12} (2010), 026

\bibitem{Ducloue:2013bva}
B.~Duclou{\'e}, L.~Szymanowski and S.~Wallon,
Phys. Rev. Lett. \textbf{112} (2014), 082003

\bibitem{Caporale:2014gpa}
F.~Caporale, D.~Y.~Ivanov, B.~Murdaca and A.~Papa,
Eur. Phys. J. C \textbf{74} (2014) no.10, 3084
[erratum: Eur. Phys. J. C \textbf{75} (2015) no.11, 535]

\bibitem{Catani:1990eg}
S.~Catani, M.~Ciafaloni and F.~Hautmann,
Nucl. Phys. B \textbf{366} (1991) 135.

\bibitem{Catani:1994sq}
S.~Catani and F.~Hautmann,
Nucl. Phys. B \textbf{427} (1994) 475

\bibitem{Schuler:1994hy}
G.~A.~Schuler,
[arXiv:hep-ph/9403387 [hep-ph]].

\bibitem{Mangano:1996kg}
M.~L.~Mangano and A.~Petrelli,
Int. J. Mod. Phys. A \textbf{12} (1997), 3887-3897

\bibitem{Feng:2015cba}
Y.~Feng, J.~P.~Lansberg and J.~X.~Wang,
Eur. Phys. J. C \textbf{75} (2015) no.7, 313



\bibitem{ColpaniSerri:2021bla}
A.~Colpani Serri, et al. 
Phys. Lett. B \textbf{835} (2022), 137556

\bibitem{Lansberg:2020ejc}
J.~P.~Lansberg and M.~A.~Ozcelik,
Eur. Phys. J. C \textbf{81} (2021) no.6, 497


\bibitem{Bartels:2001ge}
J.~Bartels, D.~Colferai and G.~P.~Vacca,
Eur. Phys. J. C \textbf{24} (2002), 83-99

\bibitem{Bartels:2002yj}
J.~Bartels, D.~Colferai and G.~P.~Vacca,
Eur. Phys. J. C \textbf{29} (2003), 235-249

\bibitem{Caporale:2011cc}
F.~Caporale, D.~Y.~Ivanov, B.~Murdaca, A.~Papa and A.~Perri,
JHEP \textbf{02} (2012), 101

\bibitem{Ivanov:2004pp}
D.~Y.~Ivanov, M.~I.~Kotsky and A.~Papa,
Eur. Phys. J. C \textbf{38} (2004), 195-213

\bibitem{Polizzi:2025edm}
A.~Polizzi, M.~Fucilla and A.~Papa,
Eur. Phys. J. C \textbf{85} (2025) no.9, 948

\bibitem{Nefedov:2019mrg}
M.~A.~Nefedov,
Nucl. Phys. B \textbf{946} (2019), 114715

\bibitem{Hentschinski:2020tbi}
M.~Hentschinski, K.~Kutak and A.~van Hameren,
Eur. Phys. J. C \textbf{81} (2021) no.2, 112
[erratum: Eur. Phys. J. C \textbf{81} (2021) no.3, 262]

\bibitem{Celiberto:2022fgx}
F.~G.~Celiberto, M.~Fucilla, D.~Y.~Ivanov, M.~M.~A.~Mohammed and A.~Papa,
JHEP \textbf{08} (2022), 092

\bibitem{Fucilla:2024cpf}
M.~Fucilla, M.~A.~Nefedov and A.~Papa,
JHEP \textbf{04} (2024), 078

\bibitem{Kniehl:2006sk}
B.~A.~Kniehl, D.~V.~Vasin and V.~A.~Saleev,
Phys. Rev. D \textbf{73}, 074022 (2006)




\bibitem{Lansberg:2021vie}
J.-P. Lansberg, M.~Nefedov and M.~A. Ozcelik,
JHEP \textbf{05} (2022) 083

\bibitem{Lansberg:2023kzf}
J.-P. Lansberg, M.~Nefedov and M.~A. Ozcelik,
Eur. Phys. J. C \textbf{84} (2024) 351

\bibitem{Flett:2024htj}
C.~A.~Flett, J.~P.~Lansberg, S.~Nabeebaccus, M.~Nefedov, P.~Sznajder and J.~Wagner,
Phys. Lett. B \textbf{859}, 139117 (2024)

\bibitem{Celiberto:2022dyf}
F.~G. Celiberto and M.~Fucilla,
Eur. Phys. J. C \textbf{82} (2022) 929

\bibitem{Fucilla:2026nlo}
M.~Fucilla, J.-P. Lansberg, M.~Nefedov, L.~Szymanowski and S.~Wallon,
JHEP \textbf{05} (2026) 164

\bibitem{Lipatov:1995pn}
L.~N. Lipatov,
Nucl. Phys. B \textbf{452} (1995) 369.

\bibitem{Nefedov:2024swu}
M.~Nefedov,
JHEP \textbf{12} (2024) 129

\bibitem{Mueller:1986ey}
A.~H. Mueller and H.~Navelet,
Nucl. Phys. B \textbf{282} (1987) 727.

\bibitem{Boussarie:2017oae}
R.~Boussarie, B.~Duclou\'e, L.~Szymanowski and S.~Wallon,
Phys. Rev. D \textbf{97} (2018) 014008

\bibitem{He:2019qqq}
Z.-G. He, B.~A. Kniehl, M.~A. Nefedov and V.~A. Saleev,
Phys. Rev. Lett. \textbf{123} (2019) 162002


\end{thebibliography}
\end{document}